\documentclass[aps,pra,twocolumn,superscriptaddress,longbibliography]{revtex4-2} 
\usepackage[version=3]{mhchem} % Formula subscripts using \ce{}
\usepackage{graphicx}  % needed for figures
\usepackage{dcolumn}   % needed for some tables
\usepackage{bm}        % for math
\usepackage{amssymb}   % for math
\usepackage{color}
\usepackage{float}

\begin{document}

\author{Harishankar Jayakumar}
\thanks{These authors contributed equally.  Current address of HJ: University of Minnesota, 321 Church St SE, Minneapolis, MN 55455, harish@umn.edu; DL: California Institute of Technology, 1200 E California Blvd, Pasadena, CA 91125, United States; BK: Rockley Photonics Inc.}
%\altaffiliation[Current address:]{Department of Physics, CUNY-City College of New York, New York, NY 10031, USA, Email: hjayakumar@ccny.cuny.edu }
%\affiliation{Institute for Quantum Science and Technology, University of Calgary, Calgary, AB, T2N 1N4, Canada}
\author{Behzad Khanaliloo}
\thanks{These authors contributed equally.  Current address of HJ: University of Minnesota, 321 Church St SE, Minneapolis, MN 55455, harish@umn.edu; DL: California Institute of Technology, 1200 E California Blvd, Pasadena, CA 91125, United States; BK: Rockley Photonics Inc.}
%\affiliation{Institute for Quantum Science and Technology, University of Calgary, Calgary, AB, T2N 1N4, Canada}
\author{David P.\ Lake}
%\affiliation{Institute for Quantum Science and Technology, University of Calgary, Calgary, AB, T2N 1N4, Canada}
\author{Paul E.\ Barclay}
%\affiliation{National Institute for Nanotechnology, Edmonton, AB, T6G 2M9, Canada}
\email{pbarclay@ucalgary.ca}
\affiliation{Institute for Quantum Science and Technology, University of Calgary, Calgary, AB, T2N 1N4, Canada}

\date{Compiled \today}
\title{Tunable amplification and cooling of a diamond resonator with a microscope}
%\title{Mechanical control of a resonator with a microscope: cooling, amplification, and mode-selective excitation}

\begin{abstract}
Controlling the dynamics of mechanical resonators is central to quantum science and metrology applications. Optomechanical control of diamond  resonators is attractive owing to diamond's excellent physical properties and its ability to host electronic spins that can be coherently coupled to mechanical motion. Using a confocal microscope, we demonstrate tunable amplification and damping of a diamond nanomechanical resonator's motion. Observation of both normal mode cooling from room temperature to 80K, and amplification into self--oscillations with  $60\,\mu\text{W}$ of optical power is observed via waveguide optomechanical readout. This system is promising for quantum spin-optomechanics, as it is predicted to enable optical control of stress-spin coupling with rates of $\sim$ 1 MHz (100 THz) to ground (excited) states of diamond nitrogen vacancy centers. 
\end{abstract}
\maketitle

\section{Introduction}

The interaction between light and mechanical systems underlies breakthroughs in physics ranging from optical tweezers \cite{ref:ashkin1970atp} to gravitational wave detection \cite{ref:scientific2017gw170104}. Nanoscale  systems  harnessing this interaction have led to  advances in quantum nanomechanics \cite{ref:groblacher2009osc, ref:chan2011lcn, ref:cohen2015pci, ref:riedinger2016ncc, ref:sudhir2017qco, ref:purdy2017qcrt, ref:riedinger2018rqe}, sensing \cite{ref:anetsberger2010mnm, ref:gavartin2012aho, ref:forstner2012com, ref:wu2017not}, and nonlinear optics \cite{ref:safavi2011eit, ref:weis2010oit, ref:dong2012odm, ref:liu2013eit, ref:lake2018oit}. An essential ingredient to many of these demonstrations is dynamic optomechanical back action, which  allows energy exchange between optical and mechanical domains \cite{ref:kippenberg2008cob, ref:aspelmeyer2014co}. Controlling diamond nanomechanical systems via optomechanical back action is of growing interest, fueled by diamond's exceptional properties \cite{ref:aharonovich2011dp}, and by demonstrations of diamond  spin manipulation using piezoelectronically driven mechanical resonators \cite{ref:lee2017trs, ref:macquarrie2013msc, ref:ovartchaiyapong2014dsc, ref:teissier2014scn, ref:barfuss2015smd, ref:ovartchaiyapong2014dsc, ref:macquarrie2015ccn, ref:meesala2016esc, ref:golter2016csa, ref:golter2016oqc, ref:maity2019cac}. Controlling  resonator motion optomechanically provides a path towards photon-phonon-spin coupling and technologies ranging from spin-spin entanglement \cite{ref:lemonde2018pns, ref:kuzyk2018pqn}  to quantum transduction \cite{ref:schuetz2015uqt}. Back action can also enhance the performance of diamond resonators used for sensing \cite{ref:rugar1991mpa, ref:barton2012pso, ref:yie2010seu}.   In this article, we show that optomechanical back action acting on modes of a diamond resonator vibrating in or out of plane can be selectively be created and controlled by adjusting the focal position of a microscope that commonly serves as an optical interface with diamond colour centres. Using this technique, we cool a diamond nanomechanical resonator, as well as amplify its motion sufficiently for mechanical control of diamond spins via their predicted coupling to phonons.

\begin{figure}[t]
\begin{center}
\includegraphics[width=1\linewidth]{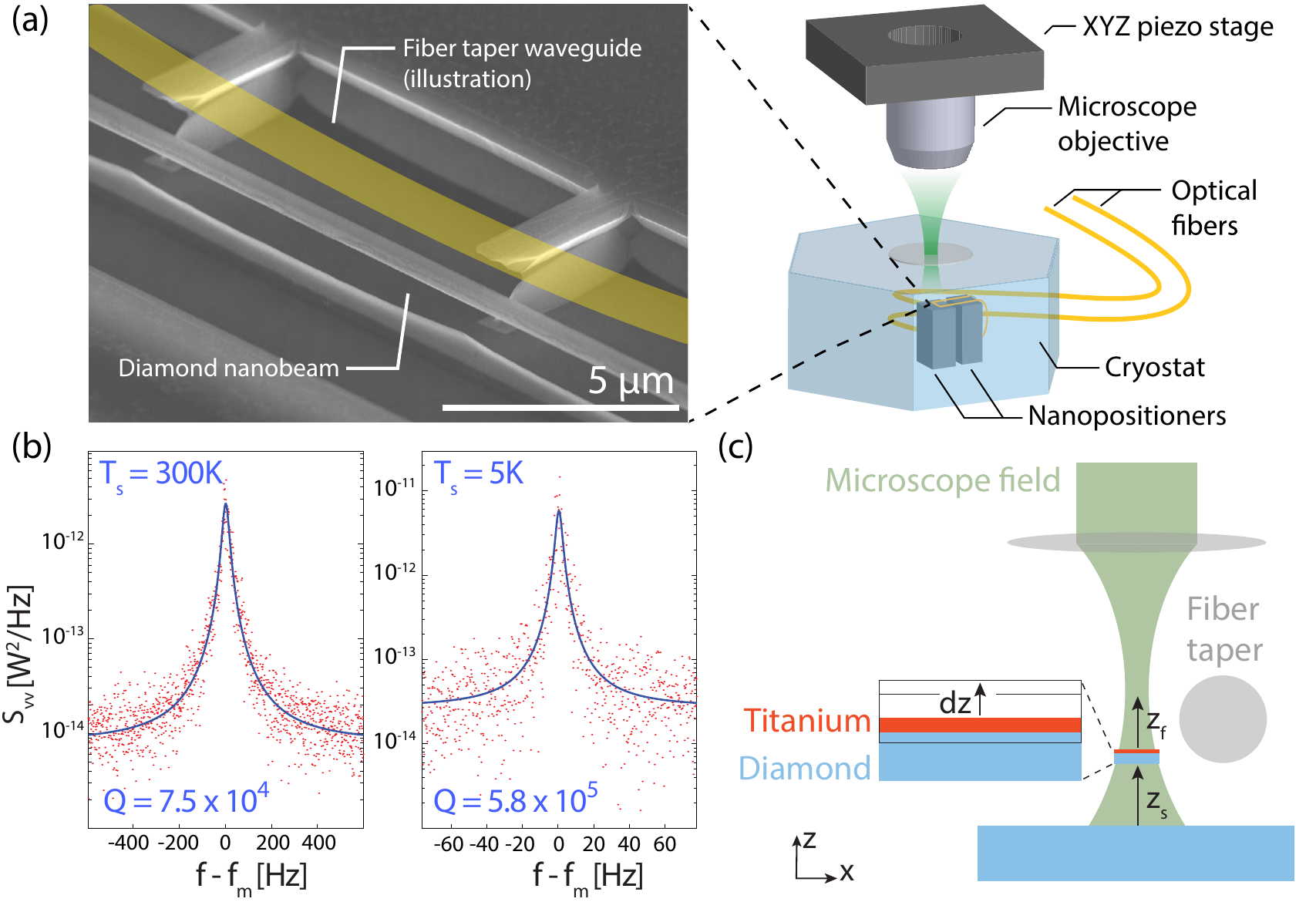}
 \caption{(a) Schematic of the optomechanical system and apparatus. A microscope objective mounted on a piezo stage focuses a green laser onto the sample. The sample and fiber taper are located in a cryostat on nanopositioners. SEM image: a diamond nanobeam similar to that used in the experiment, with an illustration of the dimpled optical fiber taper drawn in yellow in approximately the position used for evanescent coupling to the nanobeam. Also visible are diamond supports used to stabilize the fiber taper during measurements.  (b) Power spectral density of the fiber taper transmission near the $\textbf{v}_1$ nanobeam mode frequency at room and low temperature. (c) Schematic of the geometry of the optomechanical system.  }
\label{fig:schematic}
\end{center}
\end{figure}

\begin{figure*}[t]
\begin{center}
\includegraphics[width=1\linewidth]{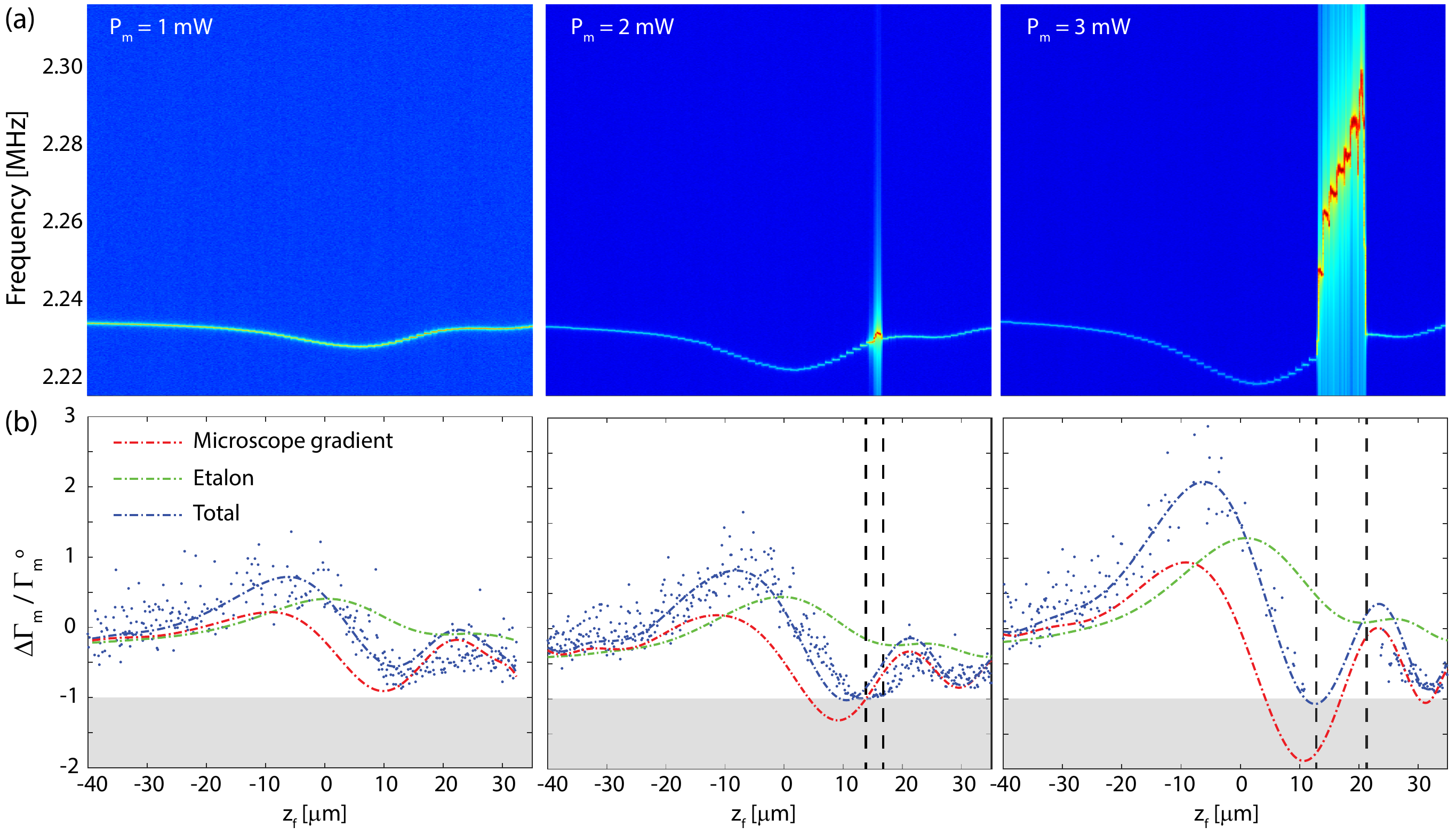}
 \caption{(a) Spectrograph showing power spectral density of the $\textbf{v}_1$ nanobeam mode as a function of microscope focus height, for varying green laser power. The sample is at room temperature. Negative $z_f$ indicates that the focal position is below the nanobeam. (b) Optomechanical  damping of $\textbf{v}_1$, normalized by its intrinsic dissipation rate, as a function of focal height and varying power. Fits are from the model in Eq.\ \eqref{eq:loss} input with a normalized $I(z_f)$ profile derived from $\Delta f_m$. Contributions from the etalon and microscope gradient terms in Eq.\ \eqref{eq:differential} are shown. Shaded regions indicate where optomechanical damping will cause self-oscillation.}
\label{fig:roomTzScan}
\end{center}
\end{figure*}

Optomechanical damping and amplification, for example by optical gradient \cite{ref:lin2009moc}, radiation pressure \cite{ ref:arcizet2006rpc, ref:gigan2006scm, ref:chan2011lcn}, or photothermal  \cite{ref:metzger2008osc, ref:metzger2004ccm, ref:favero2007ocm, ref:barton2012pso,  ref:ramos2012ows, ref:khanaliloo2015dnw} forces typically relies on feedback from a cavity \cite{ref:kippenberg2008cob, ref:aspelmeyer2014co}, waveguide coupler \cite{ref:khanaliloo2015dnw}, or external optoelectronics  \cite{ref:poggio2007fcc}.  Inspired in part by optical tweezers, here we introduce a system that operates in an optical intensity gradient dominated regime of optomechanics and does not does not require a cavity or coupling to optical resonances \cite{ref:ramos2012ows}.  The dynamic optomechanical back action is photothermal in nature, and is tuned through translation of a microscope focus, allowing both the strength and the sign of the optomechanical damping to be adjusted. This system, which adds confocal microscopy to our previously demonstrated waveguide optomechanical experiment \cite{ref:khanaliloo2015dnw},  provides a combination of  tunable optomechanical actuation and sensitive  optomechanical readout. It allows normal--mode cooling of a diamond mechanical resonator from room temperature to $\sim 80\,\text{K}$, and excitation of nanomechanical self--oscillations whose  stress field is predicted to allow control of nitrogen vacancy (NV) center spins \cite{ref:lee2017trs}. These self--oscillations are observed for continuous wave excitation from a 532 nm laser with power as low as $60~\mu\text{W}$. Unlike previous demonstrations of photothermal backaction \cite{ref:metzger2008osc, ref:metzger2004ccm, ref:favero2007ocm, ref:barton2012pso,  ref:ramos2012ows}, neither an external cavity nor a wavelength tunable laser is required to adjust the backaction  between damping and anti--damping regimes. Furthermore, we leverage its sensitivity to the microscope field gradient to selectively excite vertical as well as horizontal modes of the nanobeam, the latter of which can be difficult to probe using conventional optical interferometry measurements of nanomechanical devices \cite{Burek2013} as their motion primarily induces intensity rather than phase changes on reflected light.

\section{Device and experimental setup}

The optomechanical system studied here, illustrated in Fig.\ \ref{fig:schematic}(a),  consists of a diamond nanobeam (dimensions $l\times w\times t = 50\times0.5\times 0.25\,\mu \text{m}^3$)  illuminated by a green (532 nm) laser  input to an objective (Sumitomo long working distance, 0.55 NA) mounted on a three--axis stage. The nanobeam is fabricated from single crystal diamond (Element Six, optical grade,  3  $\times$ 3 mm$^2$ area, polished by Delaware Diamond Knives) using undercut etching  \cite{ref:khanaliloo2015dnw}, and its top surface is coated with titanium ($\sim 5$ nm thickness, deposited using electron beam evaporation), which enhances photothermal effects discussed below. The nanobeam is suspended  $\sim 2\,\mu\text{m}$ above the diamond substrate, as shown in Fig.\ \ref{fig:schematic}(a).  

In the results presented below, we show that translating the microscope controls the dynamics of the nanobeam's motion.  These dynamics are  monitored using an optical fiber taper waveguide (diameter $\sim 1\,\mu\text{m}$) \cite{ref:michael2007oft} evanescently coupled to the nanobeam, as illustrated in Fig.\ \ref{fig:schematic}(a).  The fiber taper and the diamond sample are mounted in a closed cycle cryostat (Montana Instruments) operating in high vacuum over temperatures from 5K to 300K, and are aligned using nanopositioners (Attocube). Nanobeam motion is monitored with up to $\text{fm}/\sqrt{\text{Hz}}$  sensitivity by detecting fluctuations in the coupling between the fiber taper and the nanobeam, as described in Ref.\ \cite{ref:khanaliloo2015dnw}. Nanobeam resonance dynamics are measured from the power spectral density $S_{vv}$ of the photodetected transmission of a $1570~\text{nm}$ source through the fiber taper.

Characterization of the fundamental  nanobeam vertical mechanical resonance ($\textbf{v}_1$)  in absence of the microscope field is shown in Fig.\ \ref{fig:schematic}(b), which plots $S_{vv}$ over the frequency ($f$) range spanning $\textbf{v}_1$ resonance frequency $f_m$, in high vacuum ($ < 10^{-5}$ Torr) at 300K  and 5K operating temperatures. The peak in $S_{vv}$ is thermally driven motion of $\textbf{v}_1$, whose dynamics are determined by dissipation rate $\Gamma_m = 2\pi f_m/Q_m$ where $Q_m$ is mechanical quality factor. Fitting $S_{vv}$ with a thermomechanical noise spectrum \cite{ref:cleland2002npn} we find $Q_m = 7.5\times 10^4$ and $5.8\times 10^5$ at 300K and 5K, respectively.   The fiber taper input power is sufficiently low (a few $\mu$W) so that it does not affect the dynamics.

\section{Tunable optomechanical backaction}

Turning on the microscope field introduces optomechanical back action that can be analyzed using the geometry in Fig.\,\ref{fig:schematic}(c).  The field intensity $I$ in the nanobeam depends on both the nanobeam's height above the substrate, $z_{s}$, and its distance to the microscope focal plane, $z_{f}$, and  can be approximated as $I = I_{f}(z_f)\chi(z_s)$. Here  $I_f$ describes the $z_f$ dependence of $I$. The etalon enhancement factor $\chi(z_s)$ describes interference between reflections from the etched diamond surface below the nanobeam, the titanium coated nanobeam, and the incident field, which will combine to create a standing wave pattern. In this simplified model $I_f(z_f)$ implicitly accounts for  geometry related local field corrections, for example local optical resonances of the nanobeam and the effect of the titanium layer, and we have assumed that changes in $z_s$ from nanobeam motion are sufficiently small that the etalon contribution can be treated as a separable scaling factor.  Vertical nanobeam displacement $dz$ modifies $z_{s}$ and $z_{f}$ by $\pm dz$, respectively, which in turn changes $I$. This optomechanical feedback, when combined with a lag between the nanobeam position and forces proportional to $I$,  amplifies or damps mechanical motion.

The dominant optical microscope forces on the nanobeam were found to be photothermal \cite{ref:metzger2004ccm, ref:barton2012pso}, whose optomechanical damping $\Delta\Gamma_m$ is given by 
 \begin{equation}
 \frac{\Delta\Gamma_{m}(z_f)}{\Gamma_m^o} = {Q_m^o}\frac{2\pi f_{m}^o \tau}{1+\left(2\pi f_{{m}}(z_f)\tau\right)^2}\mathcal{G}\frac{dI}{dz} \sigma,
\label{eq:loss}
\end{equation}   
where  $f_m^o$, $\Gamma_m^o$ and $Q_m^o$ are intrinsic values in absence of the microscope field. This model follows and modifies that previously analyzed in \cite{ref:khanaliloo2015dnw} in absence of a microscope field. Unlike in \cite{ref:khanaliloo2015dnw}, the field from the fiber taper does not sufficiently heat the nanobeam to induce any backaction. Instead, the nanobeam is deflected by power absorbed from the microscope field.  The nanobeam deflection for absorbed power $\sigma I$ is determined by photothermal coupling coefficient $\mathcal{G}$ (units of m/W), and depends on the nanobeam's geometry and internal compressive stress \cite{ref:khanaliloo2015dnw}. The titanium layer increases absorption cross-section $\sigma$, but is not generally necessary to observe dynamic back action  \cite{ref:khanaliloo2015dnw}. A non-instantaneous  thermal  response time $\tau$ is required for optomechanical heating or cooling. Finite element (COMSOL) simulations predict $2\pi f_m\tau\sim 3$, accounting for the reduced thermal conductivity of nanostructured diamond \cite{ref:wu2012tcd}.  

\begin{figure}[!t]
\begin{center}
\includegraphics[width=1.0\linewidth]{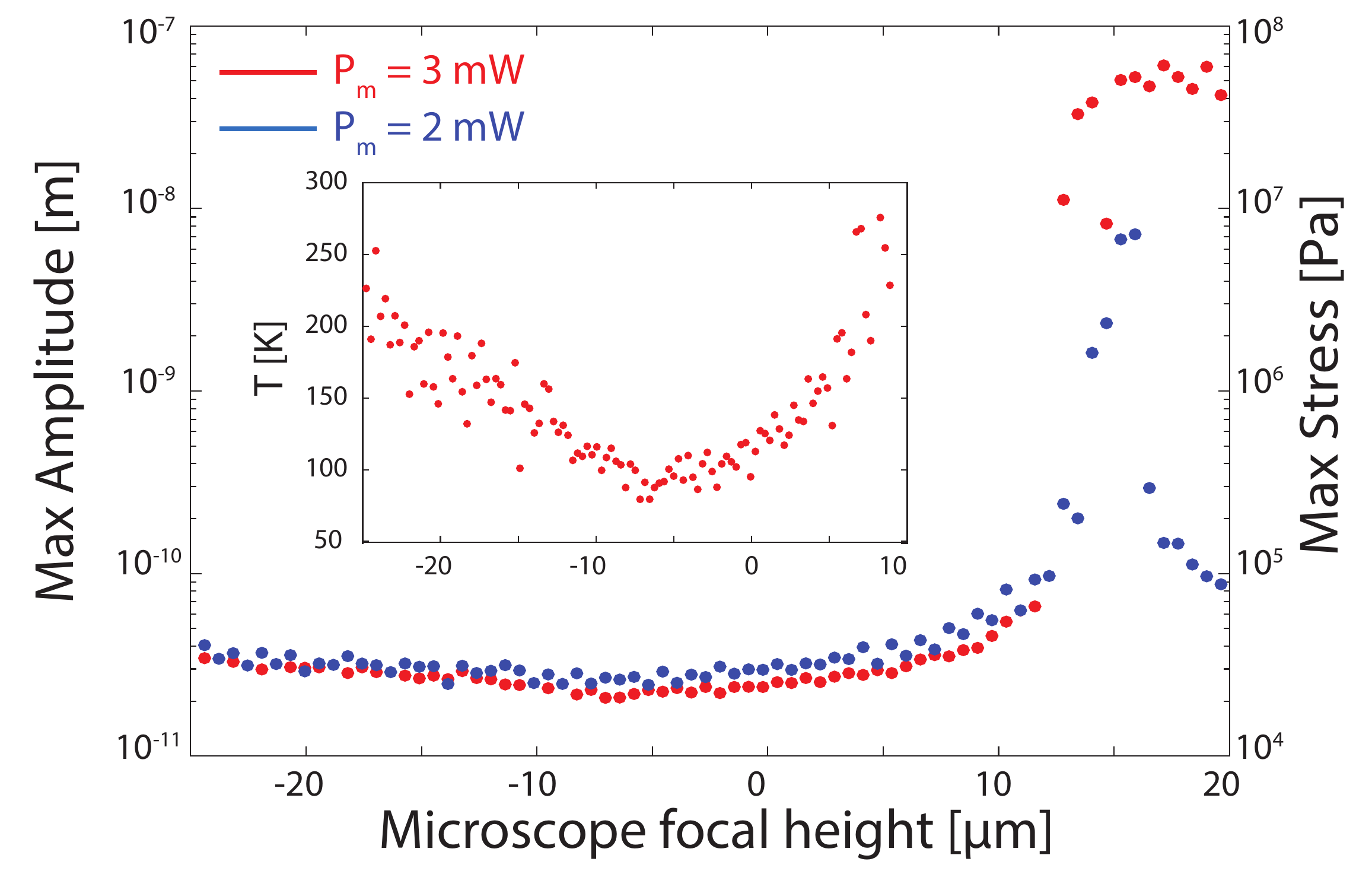}
 \caption{Amplitude (left axis) and maximum internal dynamic  stress (right axis) of the $\textbf{v}_1$ mode at room temperature as a function of microscope focal plane height for 2 mW  and 3 mW microscope power. The sample is at room temperature. When the nanobeam self oscillates, maximum stress just below $\sim$ 100 MPa near the nanobeam clamping points can be realized. Inset: Effective normal mode temperature for 3 mW microscope power as a function of microscope height scanning through the regime of maximum damping.}
\label{fig:amplitude}
\end{center}
\end{figure}

The gradient of the microscope intensity plays a critical role in determining whether $dI/dz$, and as a result $\Delta\Gamma_m$, is positive or negative. This is in contrast to cavity optomechanics, where back action is dominated by  $\chi$ whose sign is independent of external optics. To study the microscope back action, the objective was aligned with the center of the nanobeam and scanned vertically ($1~\mu\text{m}$ steps, 2.9 s/step) while monitoring $S_{vv}$.  Figure \ref{fig:roomTzScan}(a) shows this measurement at room temperature for microscope powers $P_m = 1,2$ and 3 mW. The mechanical frequency $f_m(z_f) = f_m^o + \Delta f_m(z_f)$ decreases as the microscope is focused on the nanobeam, consistent with optical heating of a compressively stressed device \cite{ref:khanaliloo2015dnw}. $\Delta f_m$ follows a profile reminiscent of the microscope laser intensity's $z_f$ dependence, providing a measure of the $I_f(z_f)$ profile that can be input to the model in Eq.\ \eqref{eq:loss}, as discussed  below. The asymmetry and oscillations in $\Delta f_m(z_f)$ are related to aberrations from the cryostat window \cite{ref:nasse2010rmi}. In general, $\Delta f_m$ is also affected by dynamic photothermal, and dynamic and static optical gradient force effects. However, they are predicted to be smaller than the observed $|\Delta f_m|$ \cite{ref:khanaliloo2015dnw}.

The influence of the microscope on the nanobeam dynamics is revealed dramatically  in Fig.\ \ref{fig:roomTzScan}(a) near $z_f = 15~\mu\text{m}$, where for $P_m = 2$ and 3 mW the peak value of $S_{vv}$ increases, indicating nanomechanical self-oscillation, and $f_m$ shifts due to nonlinear nanomechanical effects related to  large amplitude motion \cite{ref:khanaliloo2015dnw}. To analyze this quantitatively, the measured $\Delta\Gamma_m(z_f)$ is plotted in  Fig.\ \ref{fig:roomTzScan}(b), showing that motion is either damped ($\Delta\Gamma_m > 0$) or amplified ($\Delta\Gamma_m < 0$) depending on the microscope focus: the sign of $\Delta\Gamma_m$ changes as the focus is scanned from above to below the nanobeam. Near $z_f = 15\,\mu\text{m}$,  $\Gamma_m \sim 0$,  the nanobeam enters a regime of self-oscillation, in agreement with the increase in  $S_{vv}$ peak amplitude. 

This behavior illustrates a key feature of this system: the dependence of the sign of $\Delta\Gamma_{m}(z_f)$ on the microscope intensity gradient. By fitting the data in Figs.\ \ref{fig:roomTzScan}(b) with the model from Eq.\ \eqref{eq:loss}, the relative contribution from the microscope gradient and the etalon were extracted. This requires expanding the intensity gradient,
 \begin{equation}
	\frac{dI}{dz} = -\chi \left( \frac{dI_f(z_f)}{dz_{f}} - I_f(z_f)\frac{1}{\chi}\frac{d\chi}{dz_s}\right),
\label{eq:differential}
\end{equation}
and inferring $dI_f/dz_f$ and $I_f(z_f)$  from  $\Delta f_m(z_f)$ to within a proportionality constant.  In addition to this constant, the fit requires a fitting parameter $\propto \chi$ that governs the relative contributions of the intensity gradient and the etalon terms in Eq.\ \eqref{eq:differential}. Contributions from these two terms are shown in Fig.\ \ref{fig:roomTzScan}(b), showing that  in our experiment the microscope gradient is the dominant factor while the smaller etalon contribution damps mechanical motion and shifts the zero of $\Delta\Gamma_m(z_f)$. The imperfect fits reveal the approximate nature of the model. For example, it is possible that the microscope position and the etalon response, which in general is a standing wave pattern, are not  entirely separable. In future, detailed numerical simulations of the microscope field and its interaction with the nanobeam and the surrounding diamond structure would provide additional insight into optimization of the strength of the gradient contribution and minimization of the etalon contribution.

\begin{figure}[t]
\begin{center}
\includegraphics[width=1.0\linewidth]{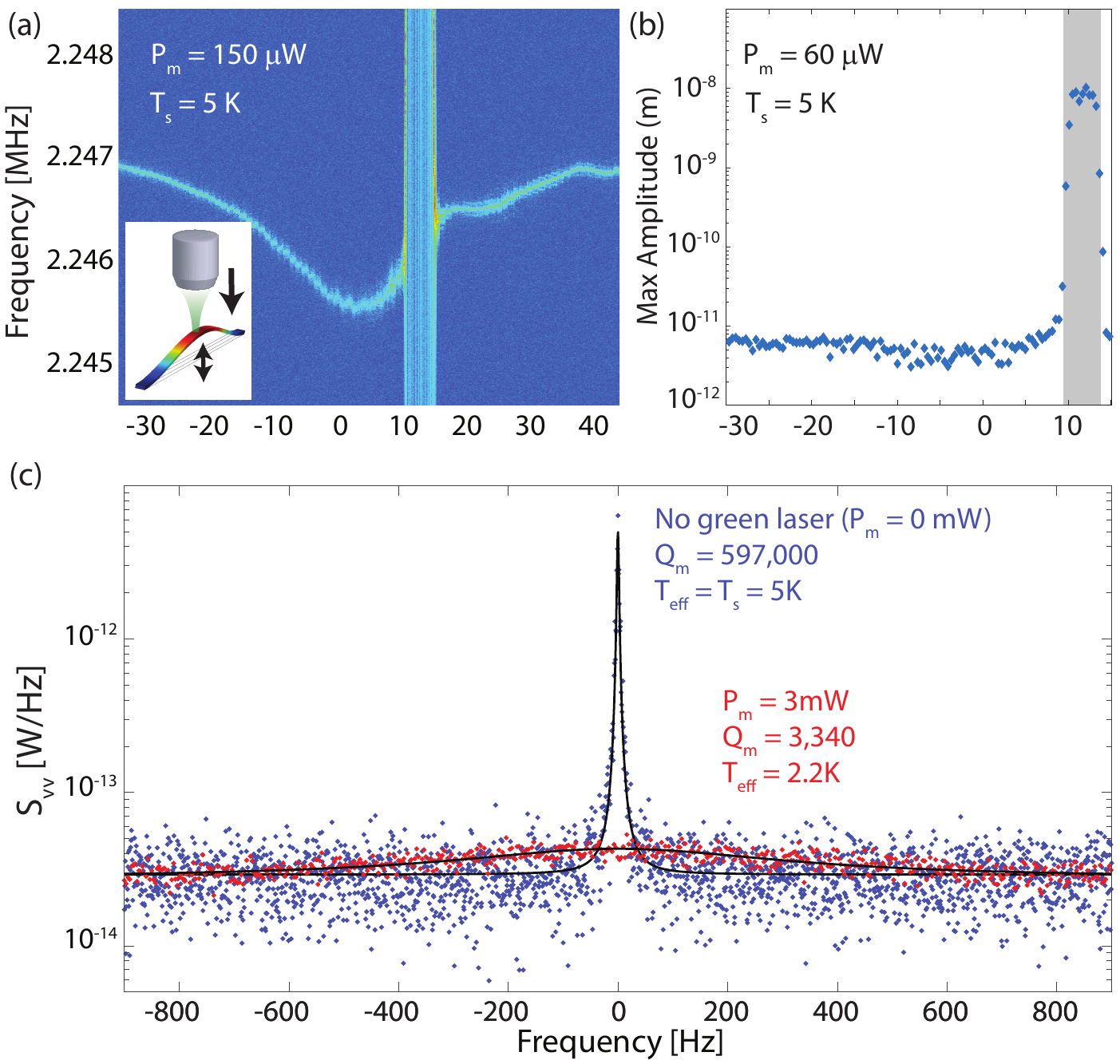}
 \caption{ Nanobeam-microscope optomechanics at cryogenic temperature $T_s = 5\,\text{K}$: (a) Spectrograph and (b) oscillation amplitude of the  $\textbf{v}_1$ nanobeam mode  motion for varying focal plane height, detected via fiber taper transmission.  (c) Power spectral density of the fiber taper transmission detected $\text{v}_1$ mode motion with the microscope off ($P_m = 0$ mW) and on ($P_m = 3$ mW) with $z_f$ optimized to maximize $\Delta\Gamma_m$.}
\label{fig:60uW}
\end{center}
\end{figure}

The amplification and damping  is further analyzed in Fig.\ \ref{fig:amplitude}, which plots the RMS amplitude for varying $z_f$ and $P_m$, extracted from the area under $S_{vv}$ normalized by the thermomechanical vibration amplitude in absence of the microscope field \cite{ref:mitchell2016scd}. When $P_m = 3~\text{mW}$, the self-oscillations reach close to 100 nm, three orders of magnitude greater than the nanobeam's intrinsic thermal motion. In contrast, when the microscope position is set to $z_f \sim -5~\mu\text{m}$,  the thermal motion of the nanobeam is damped,  cooling the resonance to $T_\text{eff}  \sim 80~\text{K}$ from the sample  temperature $T_s = 300~\text{K}$, as shown in the inset to Fig.\ \ref{fig:amplitude}. This inference of temperature from resonance area was found to be consistent with $T_\text{eff} = T_s /(1+\Delta\Gamma_m/\Gamma_m^o)$ predicted from $\Delta\Gamma_m$ in Fig.\ \ref{fig:roomTzScan}(b) \cite{ref:aspelmeyer2014co}.

\section{Cryogenic operation}

\subsection{Low power self-oscillations and cooling}

The impact of back action is increased in cryogenic conditions where the intrinsic mechanical dissipation of the diamond resonator is reduced. The improvement in device performance at low temperature is illustrated in Fig.\ \ref{fig:schematic}(b), which shows that  $Q_m^o$ increases by an order of magnitude when the sample temperature is lowered from 300 K to 5 K. As a result, for a given optomechanical backaction $|\Delta\Gamma_m|$, which is nominally independent of temperature and $Q_m^o$, the relative change in mechanical dissipation, $|\Delta\Gamma_m/\Gamma_m^o| \propto Q_m^o$, increases. This lowers the power required for optomechanical self-oscillation ($\Delta\Gamma_m/\Gamma_m^o = -1$)  or cooling to a desired $T_\text{eff}$. 

Figures\ \ref{fig:60uW}(a) and \ref{fig:60uW}(b) illustrate this effect by  showing  $S_{vv}$  and the mechanical vibration amplitude, respectively, at $T_s = 5\,\text{K}$, for varying $z_f$. In these measurements the  microscope power was reduced to $P_m = 150\,\mu\text{W}$ and $60\,\mu\text{W}$, respectively. Despite the order of magnitude lower $P_m$ compared to the room temperature measurements in Figs.\ \ref{fig:roomTzScan} and \ref{fig:amplitude}, self-oscillations with comparable amplitude are observed. 

A tantalizing prospect given the nanobeam's increase in $Q_m^o$ at cryogenic temperature is optomechanical cooling: for the $\textbf{v}_1$ mode $T_\text{eff} \approx 0.3~\text{K}$ is naively expected at $T_s = 5~\text{K}$ from the $\Delta\Gamma_m$ observed at room temperature in Fig.\ \ref{fig:roomTzScan}(b).  Figure\ \ref{fig:60uW}(c) compares $S_{vv}$ of this mode at $T_\text{eff} = 5\,\text{K}$  with and without the 3 mW microscope field turned on. With the field on and the focus optimized to maximize damping, $Q_m \sim 3340$ is inferred from the resonance linewidth. Although this is a two orders of magnitude increase in linewidth, the corresponding measured area under $S_{vv}$ was only reduced by a factor of 2.4 by the microscope field, resulting in $T_\text{eff} = 2.2~\text{K}$. 

%The discrepancy between the large linewidth broadening and the comparatively modest change in  $T_\text{eff}$ can arise from several sources.  At $T_s = 5 \text{K}$ the specific heat of diamond is four orders of magnitude smaller than at room temperature. As a result, the microscope field can more easily increase the bath temperature of the nanobeam, counteracting optomechanical damping.  In addition, evidence that others mechanisms are broadening the observed spectra can be seen by comparing the optomechanically damped linewidth at $T_{eff} = 5\,\text{K}$ and 300 K, which imply that $\Delta\Gamma_m |_\text{5\,K} \sim 2\pi f_m/5000$. This value is much larger than the room temperature measurement of $\Delta\Gamma_m |_\text{300\,K} \sim 2\pi f_m/25000$ in Fig.\ \ref{fig:roomTzScan}(b). 

The discrepancy between the large  broadening of the resonance and the comparatively modest change in  $T_\text{eff}$ can arise from several sources.  At $T_s = 5 \text{K}$ the specific heat of diamond is four orders of magnitude smaller than at room temperature. As a result, the microscope field can more easily increase the bath temperature of the nanobeam, counteracting cooling via optomechanical damping. Additional linewidth broadening could arise from fluctuations of the mechanical resonance frequency induced by the microscope field. Comparing the resonance linewidth for $P_m = 3~\text{mW}$ at $T_{s} = 5\,\text{K}$ in Fig.\ \ref{fig:60uW}(c) with the corresponding maximum room temperature  linewidth in Fig.\ \ref{fig:roomTzScan}(b), we see infer $\Delta\Gamma_m |_\text{5\,K} / \Delta\Gamma_m |_\text{300\,K} \sim 5$ if damping rate is assumed to be proportional to the linewidth. Such an enhancement in damping at low temperature requires that photothermal coupling increases in cryogenic conditions, for example due to changes in the nanobeam's compressive stress  \cite{ref:khanaliloo2015dnw}.  However,  measurements discussed below reveal that  other mechanisms can contribute significantly to linewidth broadening at cryogenic temperatures.

\subsection{In-plane mode excitation and line broadening}

\begin{figure}[t]
\begin{center}
\includegraphics[width=1.0\linewidth]{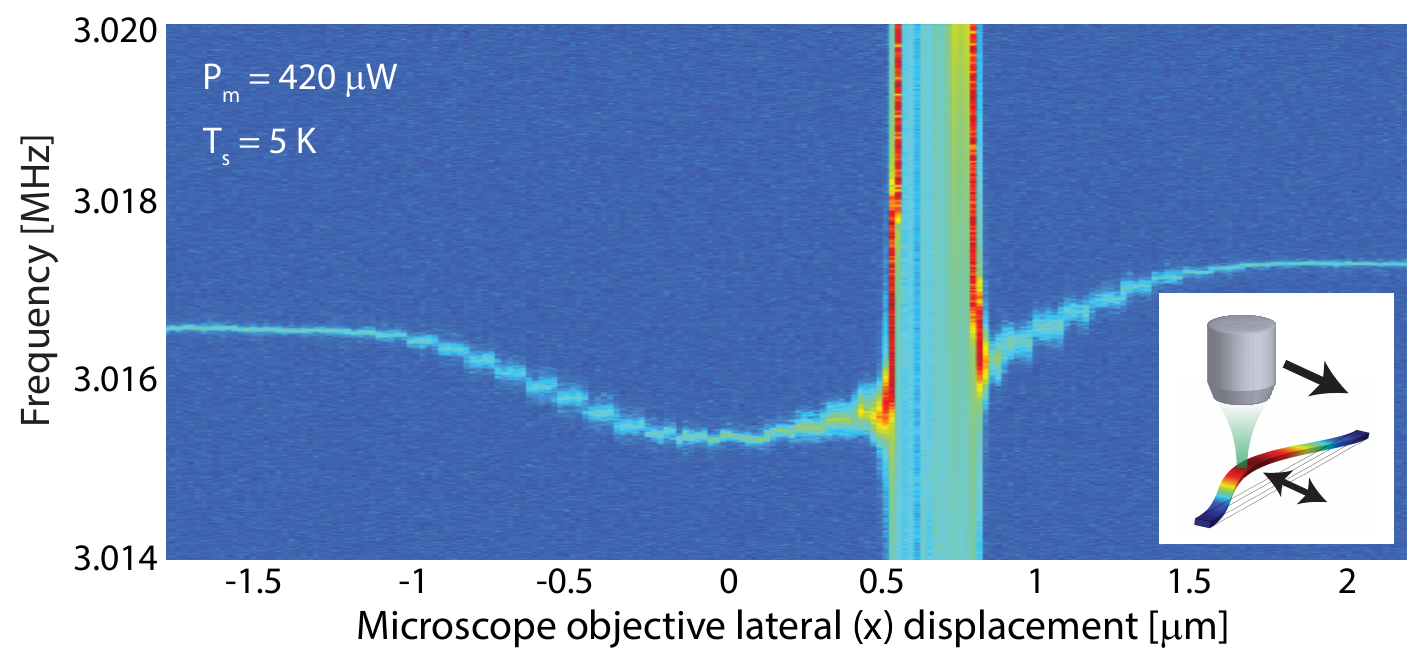}
 \caption{Exciting in-plane motion: Spectrograph of the fundamesntal in-plane $\textbf{h}_1$ nanobeam mode  motion for varying lateral focal spot position, detected via fiber taper transmission. $P_m = 420 \,\mu\text{W}$ and $T_s = 5\,\text{K}$.}
\label{fig:lateral}
\end{center}
\end{figure}

The device's higher $Q_m^o$ at cryogenic temperatures also enabled excitation of in-plane nanobeam motion. This is shown in Fig.\ \ref{fig:lateral}, which plots $S_{vv}$  of the in-plane fundamental $\textbf{h}_1$ resonance near $f_m = 3.0\,\text{MHz}$ as a function of lateral ($x$) displacement of the objective for $P_m = 450\,\mu\text{W}$.  The $4\,\mu\text{m}$  $x$ scan length is  smaller than the $z$ scans owing to the microscope's tight lateral focus in comparison to its depth of focus.  Self-oscillation occurs near $x = 0.75\,\mu\text{m}$, while for negative $x$ damping is observed. This asymmetric optomechanical response indicates that the nanobeam deflects laterally in a fixed direction when heated, independent of whether the focus is on the right or left side of the nanobeam. 

\begin{figure}[t]
\begin{center}
\includegraphics[width=1.0\linewidth]{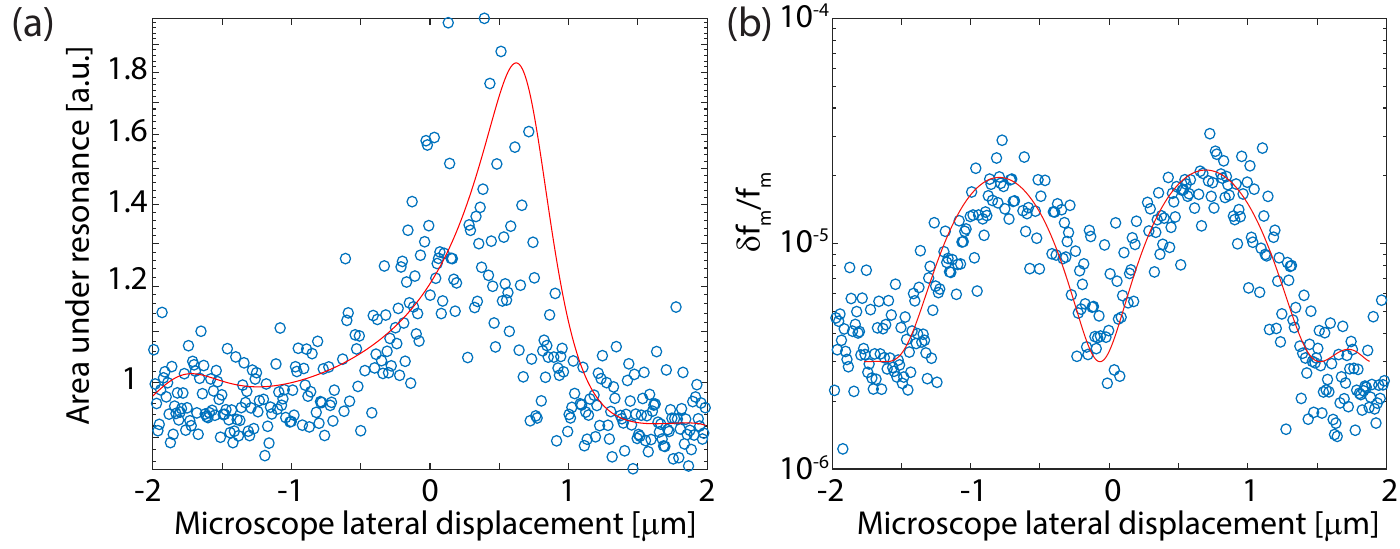}
 \caption{Competition between optomechanical backaction and thermomechanical linewidth broadening. (a) Area under the $\textbf{h}_1$ nanobeam mode spectrum  for varying lateral focal spot position. (b) Corresponding linewidth extracted by fits to the mode spectrum.  Measurements at $P_m = 200 \,\mu\text{W}$ and $T_s = 5\,\text{K}$.  }
\label{fig:broadening}
\end{center}
\end{figure}

Further analysis of the dependence of the $\textbf{h}_1$ resonance dynamics on the microscope field confirm that mechanisms in addition to optomechanical backaction broaden the mechanical lineshape.  Figure  \ref{fig:broadening}(a) plots the resonance area ($\propto T_\text{eff}$) for varying microscope displacement along $x$ with $P_m \sim 200~\mu\text{W}$ set below the threshold for self-oscillation. This clearly illustrates the asymmetric response of the optomechanical damping as a function of $x$, and is consistent with the self-oscillation data in Fig.\ \ref{fig:lateral}.  In contrast, the measured linewidth, plotted in Fig.\  \ref{fig:broadening}(b), varies symmetrically with $x$. Broadening is maximized when the nanobeam is positioned adjacent to the microscope focus, where the lateral gradient of the microscope field intensity is strong. This broadening occurs even when the nanobeam motion is being amplified. These observations suggest that  displacements of the nanobeam relative to the microscope focus are causing spectral diffusion of $f_m$.

To better understand the dependence of the mode area ($A$) and the measured linewidth ($\delta f$) on microscope position, we fit the $x$ dependent data in Fig.\ \ref{fig:broadening} with:
\begin{align}
A(x) &= A_0\left[1 + \alpha_1 \frac{I(x)}{1 + \beta \frac{dI}{dx}} \right],\\
\delta f(x) &= \delta f_o\left[ 1 + \alpha_2 \frac{dI}{dx}\right],
\end{align}
where fitting parameter $\alpha_1$ describes the increase in $T_s$ from the microscope field, and parameter $\beta$ describes the strength of the photothermal optomechanical backaction and its effect on $\Delta\Gamma$. Fitting parameter $\alpha_2$ describes a contribution to spectral broadening resulting from the nanobeam's overlap with the gradient of the microscope field intensity, for example that manifest due to variations in the position of the focal spot relative to the nanobeam. The fits in Fig.\ \ref{fig:broadening} are created by assuming that $I(x)$ is directly proportional to the measured $\Delta f_m(x)$, similar to the room temperature analysis in Fig.\ \ref{fig:roomTzScan}(b).

This model has good agreement with the data, and confirms that in addition to optomechanical damping, the linewidth is being broadened by additional mechanisms described here phenomenologically by non-zero $\alpha_2$. Further investigation into the source of this broadening is required. For example, measurements of the Allan variance of the mechanical frequency will reveal the timescale over which it is fluctuating and give insight into the nature of any technical noise affecting it. Time domain measurements of mechanical ring-down will provide a direct measurement of $\Gamma_m$. Together with power dependent measurements of $\delta f_m$, this may allow improvement of the efficiency of the photothermal cooling process in cryogenic conditions.

Note that  $T_\text{eff}$ inferred from the area of $S_{vv}$ is unaffected by spectral diffusion \cite{ref:moser2014nmr}, and that at room temperature we observe close agreement between $T_\text{eff}$ extracted from  $A$ and  $\delta f_m$, respectively. This indicates that the spectral broadening by the microscope is specific to the cryogenic measurements reported here.

\section{Discussion}

This system's potential for spin-optomechanics is significant.  Confocal microscopes are used for diamond spin spectroscopy, making this approach suited for controlling coupling of phonons and spins. For the self--oscillations reported here, dynamic stress fields of $\sim$ 100 MPa are predicted (COMSOL), as shown in Fig.\ \ref{fig:amplitude}. This corresponds to a spin-stress coupling rate $G_g/2\pi\sim$ 1 MHz and $G_e/2\pi\sim$100 THz, for the ground and excited states, respectively, of a negatively charged NV, which are comparable to coupling rates in  piezo-based stress manipulation experiments \cite{ref:lee2017trs}. Optomechanical spin control will provide a path towards creating a quantum transducer \cite{ref:schuetz2015uqt, ref:golter2016oqc} for coupling photons to spins without direct optical color center transitions, enabling interfacing telecommunication photons with spin quantum memories \cite{sh2021optomechanical}.

In conclusion, we have demonstrated optomechanical control of a diamond resonator  using a microscope. To the best of our knowledge, this is the first demonstration of tunable optomechanical damping and amplification of a nanomechanical resonator without a cavity, etalon, or other external feedback component. Using this tunable optomechanical damping, we have cooled the nanobeam's fundamental mode to below 80K, and amplified its motion sufficiently for spin-phonon coupling at rates that exceed relevant spin decoherence rates \cite{ref:lee2017trs,sh2021optomechanical}.  We have also studied the interplay between the nanomechanical resonator of the microscope field at low temperature, providing a jumping off point for future studies of  photothermal cooling in cryogenic environments.

\begin{acknowledgments}
Thank you to Aaron Hryciw, J.P. Hadden and M.\ Mitchell for assistance. This work was supported by NSERC (Discovery and Research Tools and Instruments), CFI, AITF and NRC.
\end{acknowledgments}

\end{document}